\documentclass[aps,prb,twocolumn,showpacs,groupedaddress]{revtex4}
\usepackage{graphicx}

\begin{document}

\title{The Suppression and Recovery of the Ferroelectric Phase in Multiferroic $MnWO_4$}
\author{R. P. Chaudhury$^{1}$, B. Lorenz$^{1}$, Y. Q. Wang$^{1}$, Y. Y. Sun$^{1}$, and C. W. Chu$^{1,2,3}$}
\affiliation{$^{1}$TCSUH and Department of Physics, University of Houston, Houston, TX 77204-5002} \affiliation{$^{2}$Lawrence Berkeley National
Laboratory, 1 Cyclotron Road, Berkeley, CA 94720} \affiliation{$^{3}$Hong Kong University of Science and Technology, Hong Kong, China}
\date{\today }

\begin{abstract}
We report the discovery of a complete suppression of ferroelectricity in $MnWO_4$ by 10 \% iron substitution and its restoration in external
magnetic fields. The spontaneous polarization in $Mn_{0.9}Fe_{0.1}WO_4$ arises below 12 K in external fields above 4 T. The
magnetic/ferroelectric phase diagram is constructed from the anomalies of the dielectric constant, polarization, magnetization, and heat
capacity. The observations are qualitatively described by a mean field model with competing interactions and strong anisotropy. We propose that
the magnetic field induces a non-collinear inversion symmetry breaking magnetic structure in $Mn_{0.9}Fe_{0.1}WO_4$.
\end{abstract}

 \maketitle











\section{Introduction}

Multiferroic materials in which ferroelectric (FE) and magnetic orders coexist and mutually interact have recently attracted attention because
of many novel physical phenomena observed in these compounds as well as their potential for applications as magnetoelectric sensors or memory
chips.\cite{spaldin:05,tokura:07} Common features and magnetoelectric properties of compounds with quite different chemical compositions and
structures indicate a universal physical mechanism behind the multiferroic phenomena of a large materials class. It was shown that a
non-collinear (or helical) spin density wave can break the spatial inversion symmetry and, with sufficiently strong spin-lattice coupling, can
result in a lattice distortion with a FE polarization.\cite{mostovoy:06} The helical magnetic order was in fact observed in the FE phases of
$TbMnO_3$,\cite{kenzelmann:05} $Ni_3V_2O_8$,\cite{lawes:05} $MnWO_4$,\cite{taniguchi:06} and others. Most multiferroics exhibit highly
frustrated magnetic orders due to geometric constraints or competing interactions. Local lattice distortions can therefore release the magnetic
frustration, lower the magnetic energy, and induce local electrical dipoles that may add up to a macroscopic polarization if permitted by
symmetry.\cite{delacruz:06} In frustrated magnetic systems several magnetic states are close in energy and compete for the ground state. As a
result, in multiferroic systems the FE polarization, which is induced by magnetic order, is easily controlled by small perturbations affecting
the microscopic exchange constants or the magnetic order. For example, is was shown that magnetic
fields\cite{higashiyama:04,taniguchi:06,hur:04b,delacruz:06c} and external pressure\cite{chaudhury:07,delacruz:07,chaudhury:07b} have a
significant effect on the feroelectric polarization and the multiferroic properties.

The multiferroic $MnWO_4$ passes through three magnetic transitions upon decreasing temperature $T$. The sinusoidal AF3 phase (12.6 K $<T<$ 13.5
K) is followed by a non collinear helical spin phase AF2 (7.8 K $<T<$ 12.6 K), both phases with an incommensurate (IC) modulation vector
$\overrightarrow{q_{2,3}}=(-0.214,1/2,0.457)$. The AF1 phase below 7.8 K is collinear ($\overrightarrow{q_1}=(\pm1/4,1/2,1/2)$) and shows the
typical $\uparrow\uparrow\downarrow\downarrow$ (E-type) spin pattern \cite{lautenschlager:93}. Ferroelectricity was only observed in the helical
AF2 phase,\cite{taniguchi:06,arkenbout:06} but Heyer et al. reported a finite value of the FE polarization also in the low temperature AF1
commensurate phase.\cite{heyer:06} However, our recent polarization measurements of $MnWO_4$ at ambient and high pressures\cite{chaudhury:07}
are consistent with the paraelectric nature of the AF1 phase. Magnetic frustration due to competing exchange interactions and a strong uniaxial
anisotropy lead to the sequence of complex magnetic structures. Tuning the magnetic exchange interactions appears to be imperative to arrive at
a deeper understanding of the multiferroic properties of $MnWO_4$ and related compounds. The superexchange interactions depend on the type of
magnetic ions, their interatomic distances and bond angles. The remarkable sensitivity of the FE and magnetic phases of multiferroic materials
with respect to hydrostatic pressure was demonstrated for $MnWO_4$ \cite{chaudhury:07b}, $Ni_3V_2O_8$ \cite{chaudhury:07}, and $RMn_2O_5$
\cite{delacruz:07} proving the high susceptibility of multiferroics to small perturbations. Alternatively, magnetic exchange interactions can be
controlled by replacing one kind of magnetic ions with other ions, either nonmagnetic or with different magnetic properties. In $MnWO_4$ the
magnetic $Mn^{2+}$ ion with spin S=5/2 can easily be replaced by the $Fe^{2+}$ ion with a smaller magnetic moment (S=2) and different exchange
and anisotropy constants. In an attempt to examine the detailed interplay between magnetic interactions and ferroelectricity we have therefore
investigated the solid solution $Mn_{1-x}Fe_xWO_4$. We found that $Mn_{0.9}Fe_{0.1}WO_4$ is paraelectric (PE) in zero magnetic field at all
temperatures, however, ferroelectricity and multiferroic properties are induced by magnetic fields above 4 T.

\section{Methodology}

Polycrystalline $Mn_{0.9}Fe_{0.1}WO_4$ was synthesized by solid state reaction at 930$^\circ$C of the precursor compounds $Mn_2O_3$, $WO_3$, and
$Fe_2O_3$, mixed in the appropriate ratios. Single crystals have been grown from the polycrystalline feed rod in a Floating Zone Furnace. X-ray
analysis showed the monoclinic structure with lattice constants $a$=4.799(2) {\AA}, $b$=5.736(2) {\AA}, and $c$=4.980(2) {\AA}, in good
agreement with previous reports on natural and synthetic Wolframites.\cite{guillen:85,garciamatres:03} The crystals have been characterized and
oriented by single crystal Laue diffractometry. The homogeneity of the crystals and their $Fe$ content were verified by wavelength dispersive
spectroscopy analysis. Magnetic measurements were conducted in a superconducting quantum interference device (SQUID) in magnetic fields up to 5
T. For the dielectric constant and polarization measurements a home-made dielectric probe was adapted to the Physical Property Measurement
System (PPMS) for temperature and magnetic field ($H<$ 7 T) control. The crystals were shaped as thin parallel plates, the dielectric constant
was calculated from the capacitance measured by the high-precision AH2500A capacitance bridge, and the polarization was determined by
integrating the pyroelectric current measured by a K6517A electrometer. During the pyroelectric measurements a small poling electric field was
applied in order to align domains in the ferroelectric state. The PPMS was also used for heat capacity measurements at temperatures above 1.8 K
in different magnetic fields.

\section{Experimental Results}

The low-field magnetization data are consistent with previous reports \cite{stuesser:01} and they reveal two successive phase transitions at
$T_N$=15.4 K and at $T_L$=12 K, respectively, in excellent agreement with the phase diagram of $Mn_{1-x}Fe_xWO_4$ for x=0.1 previously derived
from neutron scattering experiments \cite{garciamatres:03}. According to the neutron data magnetic order sets in at $T_{N}$ with an IC
modulation vector $\overrightarrow{q_3}=(-0.235,1/2,0.49)$, similar to the AF3 phase of pure $MnWO_4$. At $T_L$ the magnetic modulation locks
into a commensurate order described by $\overrightarrow{q_1}=(\pm1/4,1/2,1/2)$ (AF1 phase of $MnWO_4$). The spins are oriented along the easy
axis in the $ac$-plane at an angle of 35$^o$ with the $a$-axis.

\subsection{Dielectric constant and polarization}

In order to search for the signature of ferroelectricity we have performed measurements of the dielectric constant $\varepsilon(T)$ and the
pyroelectric current along all three crystallographic orientations. In zero magnetic field, we did not find any indication of a FE transition,
in contrast to the ferroelectricity observed in $MnWO_4$ \cite{taniguchi:06}. Minute changes of slope of $\varepsilon(T, H=0)$ near $T_N$ and
$T_L$ are barely visible in the inset of Fig. 1. The replacement of 10 \% $Mn$ by $Fe$ apparently leads to the complete loss of ferroelectricity
and, presumably, the loss of the helical magnetic structure in the compound. This conclusion is supported by the results of neutron scattering
experiments that did not show an intermediate phase between $T_N$ and $T_L$ in zero magnetic field.

In external magnetic fields, $H_e$, oriented along the magnetic easy axis, however, new anomalies in the dielectric properties, measured along
the monoclinic $b$-axis, could be observed. In the following we discuss dielectric and polarization data measured along the $b$-axis in external
fields $H_e$. With increasing magnetic field a distinct peak of $\varepsilon(T)$ develops near the lock-in transition temperature, $T_L$. The
peak appears for fields exceeding about 4 T and increases with $H_e$ (inset of Fig. 1). The pyroelectric current also exhibits a sharp peak
indicating the onset of FE order at $T_L$. The results for the polarization $P(T)$ at different magnetic fields are shown in Fig. 1. At 4.5 T,
close to $T_L$, $P(T)$ is small but it increases significantly at about 6 K indicating a major change within the ferroelectric phase from a
low-polarization (LP) to a high-polarization (HP) state. Similar transitions within the ferroelectric phase have been observed in other
multiferroic compounds and are sometimes attributed to spin re-orientations associated with a change of the FE polarization.\cite{delacruz:06}
The $P(T)$ data also reveal a large temperature hysteresis as shown for $H_e$=5.5 T in Fig. 1. With increasing magnetic field $P(T)$ increases
quickly and at 7 T the FE polarization increases continuously from $T_L$ towards lower $T$. From the $\varepsilon(T)$- and $P(T)$-data shown in
Fig. 1 we conclude that $Mn_{0.9}Fe_{0.1}WO_4$ is paraelectric at zero magnetic field but ferroelectricity is induced by fields above 4 T.

In order to further study the field-induced FE phase we have conducted field-dependent isothermal polarization measurements. The results (Fig.
2) unambiguously prove the ferroelectricity arising above a critical magnetic field. The transition into the FE phase exhibits a field
hysteresis of more than 0.5 T (inset of Fig. 2) confirming the first-order nature of this phase transition. $P(H)$ shows a characteristic kink
which is consistent with the sudden increase of $P(T)$ from the LP to the HP state, as discussed above. The origin of this distinct feature is
not clear and has yet to be investigated. From these data a field-temperature phase diagram for $Mn_{0.9}Fe_{0.1}WO_4$ is constructed and shown
in Fig. 3 (the hysteresis across the ferroelectric transition is shown as the dashed area). The field dependence of the N\'{e}el temperature
($T_N$) and the lock-in temperature ($T_L$) are derived from magnetic and heat capacity measurements (discussed below). Both temperatures
decrease slightly with $H_e$ as expected for an AFM state. At low fields, $T_N$ and $T_L$ define the transitions PM/PE $\rightarrow$ AF3/PE
$\rightarrow$ AF1/PE. The field-induced FE phase is labeled as AF2/FE.

\subsection{Magnetic properties}

The ferroelectricity in most multiferroic compounds is strongly coupled to the magnetic order. Therefore, any change of the dielectric state
should also be detected as anomalies of the magnetic properties. The magnetization measurements of $Mn_{0.9}Fe_{0.1}WO_4$ shown in Fig. 4 indeed
reflect the major phase transitions observed in dielectric and polarization measurements. At the N\`{e}el temperature $M(T)$ exhibits the
characteristic maximum with the onset of IC AFM order. $M(T)$ shows a subtle change with a temperature hysteresis of about 2 K at $T_L$. This
hysteresis indicates the first order nature of the lock-in transition. In high magnetic fields and at low temperature $M(T)$ reveals another
distinct anomaly with a wide hysteresis similar to the polarization data shown in Fig. 1. The change of the magnetic order at the high-field FE
transition is also detected in the $M(H)$ measurements (inset of Fig. 4). The sharp increase of $M(H)$ above 4 T and the observed hysteresis
suggests that the magnetic order undergoes a major modification. Note that the maximum field of the magnetometer (5 T) is not enough to complete
the transition into the high-field phase, however, the critical fields and temperatures derived from the decreasing field cycle (triangles in
Fig. 3) agree well with the similar data from the $P(H)$ measurements (squares in Fig. 3). We conclude that the magnetic field along the easy
axis induces a new magnetic order that breaks the spatial inversion symmetry and is compatible with the FE order. Any other direction of the
magnetic field was found to be less or not at all efficient in stabilizing the FE phase. We have also extended the pyroelectric measurements
along the $a$- and $c$-axes at zero and high magnetic fields but we did not find any signature of FE order along these orientations.

\subsection{Heat capacity}

The thermodynamic signature of transitions between different phases or orders is usually detected in distinct anomalies of the heat capacity,
$C_p(T)$. Multiferroic materials with a sequence of subsequent transitions may show more or less pronounced sudden changes of $C_p$, some can be
subtle and difficult to detect.\cite{delacruz:06} Fig. 5 shows the heat capacity data, $C_p/T$, for $Mn_{0.9}Fe_{0.1}WO_4$ in different external
magnetic fields $H_e$. For $H_e$=0 a sharp increase of $C_p(T)$ at $T_N$ indicates the onset of the IC sinusoidal order. At the lock-in
transition $C_P$ exhibits a sharp peak which is difficult to resolve in detail because of the strong first order nature of this transition. No
significant field dependence of $C_p/T$ is noted up to 3 T. The only change in this low-field range is a small shift of the step-like increase
at $T_N$ and the peak at $T_L$ towards lower $T$. At higher magnetic fields, however, $C_p/T$ shows a well resolved enhancement below the
critical temperature of the ferroelectric order (indicated by dashed arrows in Fig. 5). This anomaly quickly shifts to higher $T$ with
increasing $H_e$ and it merges with the lock-in temperature, $T_L$, at the highest field (7 T in Fig. 5). The enhanced heat capacity is a clear
thermodynamic signature of the ferroelectric order in the AF2/FE phase and it proves the bulk nature of the FE high-field phase.

\section{Model calculations}

Ferroelectricity induced by a magnetic field is a rare phenomenon and it shows the sensitivity of the frustrated magnetic system towards small
perturbations. A simple model describing the different magnetic structures (commensurate $q=1/4$ at $T=0$, IC helical and sinusoidal orders at
$T>0$) has to include the competing exchange interactions, the magnetic anisotropy, and the external field ($\overrightarrow{H}$). It should be
noted that the spin order in the low-temperature commensurate phase of $MnWO_4$ can be represented by a stack of spin chains in the $ac$ plane
with equal spin values within each chain but a fourfold modulation ($q=1/4$) of moments perpendicular to the chains. The order of spin chains is
then described by the characteristic $\uparrow\uparrow\downarrow\downarrow$ (E-type) modulation.\cite{lautenschlager:93} Within a simplified
description we can describe the major features of the magnetic order by a mean field model of spin chains stacked along one dimension with
effective exchange coupling and anisotropy parameters.

The Heisenberg model with competing nearest ($J_1$) and next nearest neighbor interactions ($J_2$) and spin anisotropy ($K$) interpolates
between the helical ground state (isotropic model) \cite{nagamiya:67} and the E-type ($\uparrow\uparrow\downarrow\downarrow$) modulation for
strong uniaxial anisotropy (Ising limit).\cite{bak:82} A similar model has been employed to describe qualitatively the magnetic phase sequence
in $Ni_3V_2O_8$.\cite{kenzelmann:06} For simplicity we consider the S=1 model with the easy axis of magnetization along the $z$-direction
\begin{eqnarray}
H=J_1\sum \overrightarrow{S}_n\overrightarrow{S}_{n+1}+J_2\sum \overrightarrow{S}_n\overrightarrow{S}_{n+2}\nonumber\\
-K\sum(S_n^z)^2-\overrightarrow{H}\sum \overrightarrow{S}_n
\end{eqnarray}
which includes the most basic parameters and interactions to describe the various spin orders in $Mn_{1-x}Fe_xWO_4$ and other multiferroics.
Neutron scattering experiments on $MnWO_4$ suggest a competition between ferromagnetic (FM, $J_1<0$) and AFM ($J_2>0$) exchange interactions and
a strong uniaxial anisotropy ($K>0$) favoring the spin orientation along the easy axis in the $ac$-plane.\cite{lautenschlager:93} The model (1)
is solved in the mean field approximation. In this approximation the two-spin interaction is replaced by a single spin term in an effective
field of the neighboring spins that is proportional to the corresponding exchange coupling constants and the thermal average of the spins at
nearest or next-nearest neighbor sites. The thermodynamic average of all spins is then calculated selfconsistently from a set of N coupled
equations, with N being the total number of spins along the chain. For simple commensurate spin orders the set of N equations can be reduced to
a smaller number of equations assuming translational symmetry of the average spin values according to the specific order. For example, for a
two-sublattice antiferromagnetic order the selfconsistency equations are reduced to two coupled equations, one for each sublattice. However, in
the case of arbitrary (commensurate or incommensurate) magnetic orders no assumption about the periodicity of the magnetic order can be made and
a large set of N coupled equations has to be considered. We solve the mean-field equations for finite lattices of up to 100 spins assuming
periodic boundary conditions. The solution of the mean-field equations for a given set of parameters ($K$, $J_1$, $J_2$, $T$, or magnetic field)
without any initial constraints on the magnetic order reproduces the different magnetic structures observed in $Mn_{1-x}Fe_xWO_4$ and $MnWO_4$,
E-type and sinusoidal (IC) collinear as well as IC helical spin modulations. Wherever more than one type of order was obtained selfconsistently
from the equations the solution with the lower free energy was determined as the stable magnetic configuration.

The ground state phase diagram of (1) for zero field is shown in Fig. 6. At $K=0$ the ground state is FM for $J_2/|J_1|<1/4$ and it assumes the
non-collinear (NC) incommensurate spin structure for larger $J_2$. For large K the Ising limit is reproduced and the transition from FM to the
collinear E-type magnetic order takes place at $J_2/|J_1|=1/2$ as in the ANNNI model.\cite{bak:82} The E-type phase is stable for larger $J_2$
and $K$ and the NC helical phase covers the left section of the ground state phase diagram. For $J_2/|J_1|\approx 1/2$ the helical spin
component perpendicular to the easy axis rapidly decreases with increasing $K$ and it disappears at the transition into the collinear sinusoidal
structure close to $K/|J_1|\approx 1.25$ (vertically dashed area in Fig. 6). The horizontally dashed area in Fig. 6 marks the region of special
interest. Within this area the NC helical spin order appears at finite temperatures between the E-type ground-state and the sinusoidal phase.
This phase sequence is typical for $MnWO_4$. With increasing $K$, however, the helical phase becomes unstable at all $T$ and, with increasing
temperature, the transition proceeds directly from the E-type state to the sinusoidal phase, as observed for $Mn_{0.9}Fe_{0.1}WO_4$. The lack of
ferroelectricity (at zero field) in $Mn_{0.9}Fe_{0.1}WO_4$ is therefore explained by the increase of the uniaxial anisotropy $K$. Evidence for
the increase of $K$ with $Fe$-substitution was also derived from the results of neutron scattering
experiments.\cite{stuesser:01,garciamatres:03} The chosen set of parameters, $J_2\approx |J_1|$, results in a spin modulation in the IC phase,
$Q_{SIN}=0.21$, that is comparable with the experimental results.\cite{nagamiya:67,lautenschlager:93}

In order to understand the magnetic field induced ferroelectricity in $Mn_{0.9}Fe_{0.1}WO_4$ we have to consider the model (1) in fields
oriented along the easy axis ($z$-axis) and search for NC (helical) spin structures. The solution of (1) in the mean field approximation for
parameters $J_1=-1$, $J_2=1$, $K=1.05$, and $\overrightarrow{H}=(0,0,H^z)$ indeed reveals a magnetic field induced transition from the E-type
order to a NC helical phase at low temperatures. The $H^z-T$ phase diagram for this case is shown in Fig. 7a. At zero magnetic field the phase
sequence with increasing $T$ is E-type $\Rightarrow$ SIN $\Rightarrow$ PM but above a critical field the helical spin structure is stabilized as
an intermediate phase. The helical and the sinusoidal phases are incommensurate with a modulation comparable to the experimental results. It is
remarkable that the simple model (1) describes the main physical effects of the $Fe$-substitution, namely: (i) the loss of the FE (helical)
phase with the increase of the anisotropy ($Fe$-substitution) and (ii) the re-appearance of this phase above a critical magnetic field oriented
along the easy axis of magnetization. For lower anisotropy ($K=0.8$) the known magnetic phase diagram \cite{arkenbout:06} for $MnWO_4$ is
qualitatively reproduced, as shown in Fig. 7b. Note that the sinusoidal spin order is also suppressed by $H^z$ in agreement with the
experimental results for $MnWO_4$. The phase transitions from the E-type to sinusoidal and helical phases (Fig. 7) are first order transitions
whereas a second order transition takes place from the sinusoidal to the paramagnetic phase, in accordance with our experimental results
discussed above.

\section{Summary and Conclusions}

We have shown that the ferroelectric and helical magnetic orders in $MnWO_4$ are easily suppressed by replacing only 10 \% of the magnetic
$Mn^{2+}$ ion by another magnetic ion, $Fe^{2+}$. The ferroelectric state is restored in external magnetic fields applied along the easy axis of
the spins. We have resolved the field-temperature phase diagram for $Mn_{0.9}Fe_{0.1}WO_4$ showing three distinct magnetic and/or ferroelectric
phases. The experimental observations are explained by a mean field solution of the Heisenberg model with competing interactions, uniaxial
anisotropy, and external magnetic field. The destruction of the ferroelectric/helical phase is due to an increase of the uniaxial anisotropy
with the iron substitution. The calculated phase diagrams describe the experimental results for $Mn_{0.9}Fe_{0.1}WO_4$ and $MnWO_4$
qualitatively well. While the model is too simple to allow for any quantitative comparison with the experiments it includes the major physical
interactions to describe qualitatively the different magnetic phases in $MnWO_4$ and $Mn_{0.9}Fe_{0.1}WO_4$. Improvements of the model
description should include the 3d lattice structure and spin order, the differences in the exchange coupling constants between $Mn$- and
$Fe$-ions, and the effects of disorder. The results of our simple model calculation should also be relevant to other multiferroic materials such
as $CuFeO_2$ where magnetic field induced ferroelectricity has been observed very recently.\cite{kimura:06b,kanetsuki:07} Our experimental
results and calculations for $Mn_{0.9}Fe_{0.1}WO_4$ suggest that with increasing magnetic field the collinear E-type spin modulation turns into
a helical IC magnetic order breaking the spatial inversion symmetry which allows for the observed ferroelectricity. This transition could be
verified by high-field neutron scattering experiments. It is expected that these experiments also reveal additional details of the magnetic
orders within the FE phase as observed by us in $Mn_{0.9}Fe_{0.1}WO_4$. Large single crystals are currently being grown for further
investigations.

\begin{acknowledgments}
This work is supported in part by the T.L.L. Temple Foundation, the J. J. and R. Moores Endowment, and the State of Texas through TCSUH and at
LBNL through the US DOE, Contract No. DE-AC03-76SF00098.
\end{acknowledgments}

\bibliographystyle{phpf}


\begin{figure}
\caption{(Color online) Polarization $P(T)$ of $Mn_{0.9}Fe_{0.1}WO_4$ in different fields above 4 T (cooling data included at 5.5 T as dashed
line, all other data are shown for increasing $T$). Inset: $\varepsilon(T)$ at fields from 0 to 7 T (curves are vertically offset).}
\caption{(Color online) Isothermal field dependence of the FE polarization (data shown for decreasing field only). Inset: 5 K data with
increasing an decreasing field.}
\caption{(Color online) Magnetic phase diagram of $Mn_{0.9}Fe_{0.1}WO_4$ with H along the easy axis. The FE phase is labeled AF2/FE. Data are
from polarization (squares) and magnetization (triangles) measurements. $T_{FE}$ was determined from $P(H)$ and $M(H)$. The shaded area
indicates the field hysteresis of the FE transition.}
\caption{(Color online) High-field magnetization $M$ of $Mn_{0.9}Fe_{0.1}WO_4$. The loop at low $T$ indicates the transition into the AF2/HP
phase. Inset: $M(H_e)$ at different temperatures (data for $T<$5 K are vertically shifted).}
\caption{(Color online) Heat capacity $C_p/T$ of $M$ of $Mn_{0.9}Fe_{0.1}WO_4$ at zero and high magnetic fields (field along the easy axis). The
dashed vertical arrows point to the anomaly associated with the ferroelectric transition at 5, 6, and 7 T, respectively (left to right).}
\caption{(Color online) Ground state phase diagram of the model (1) in mean field approximation.}
\caption{Field-Temperature phase diagrams for model parameters qualitatively describing the phase sequence in (a) $Mn_{0.9}Fe_{0.1}WO_4$ and (b)
$MnWO_4$.}
\end{figure}


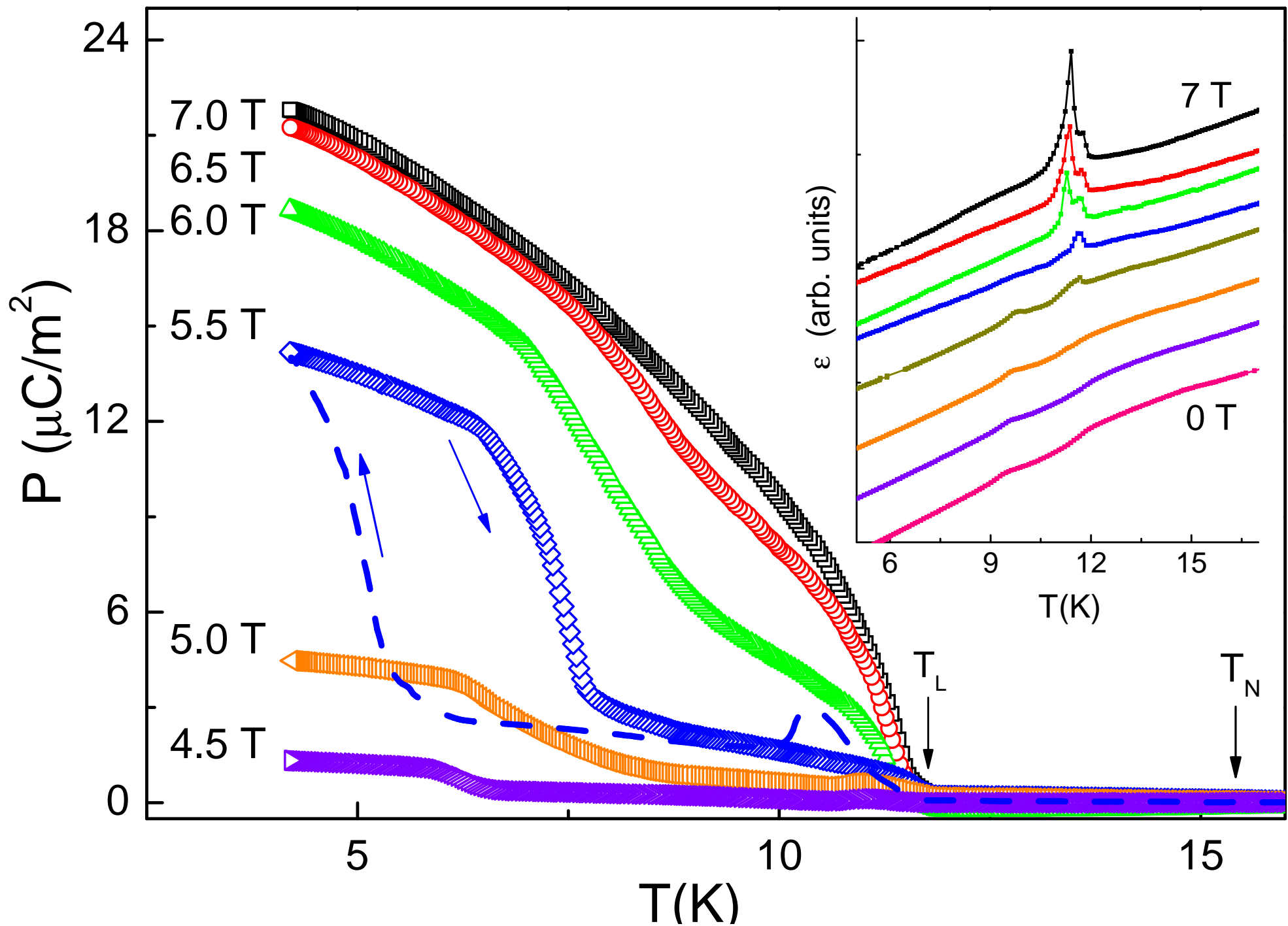
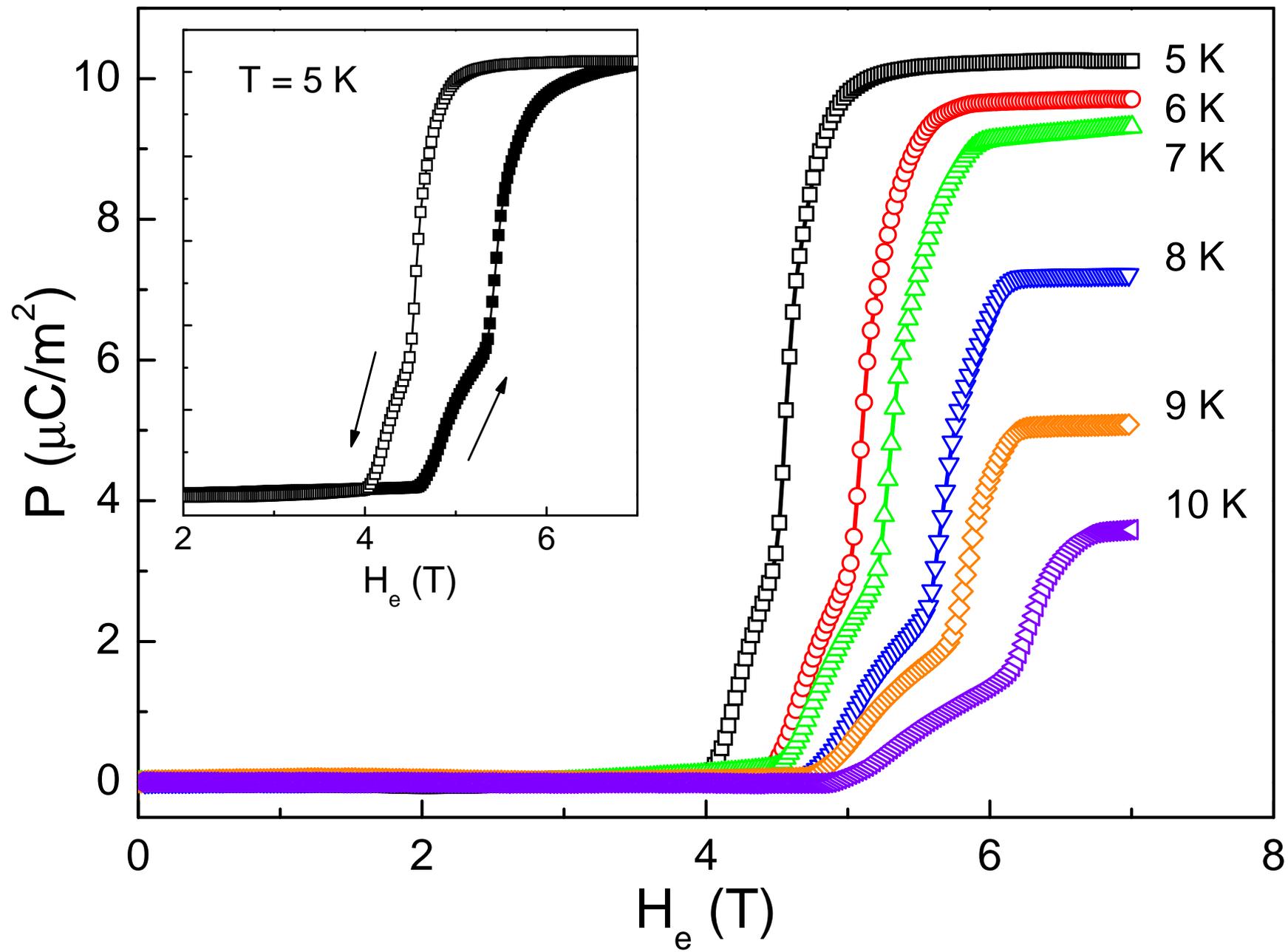
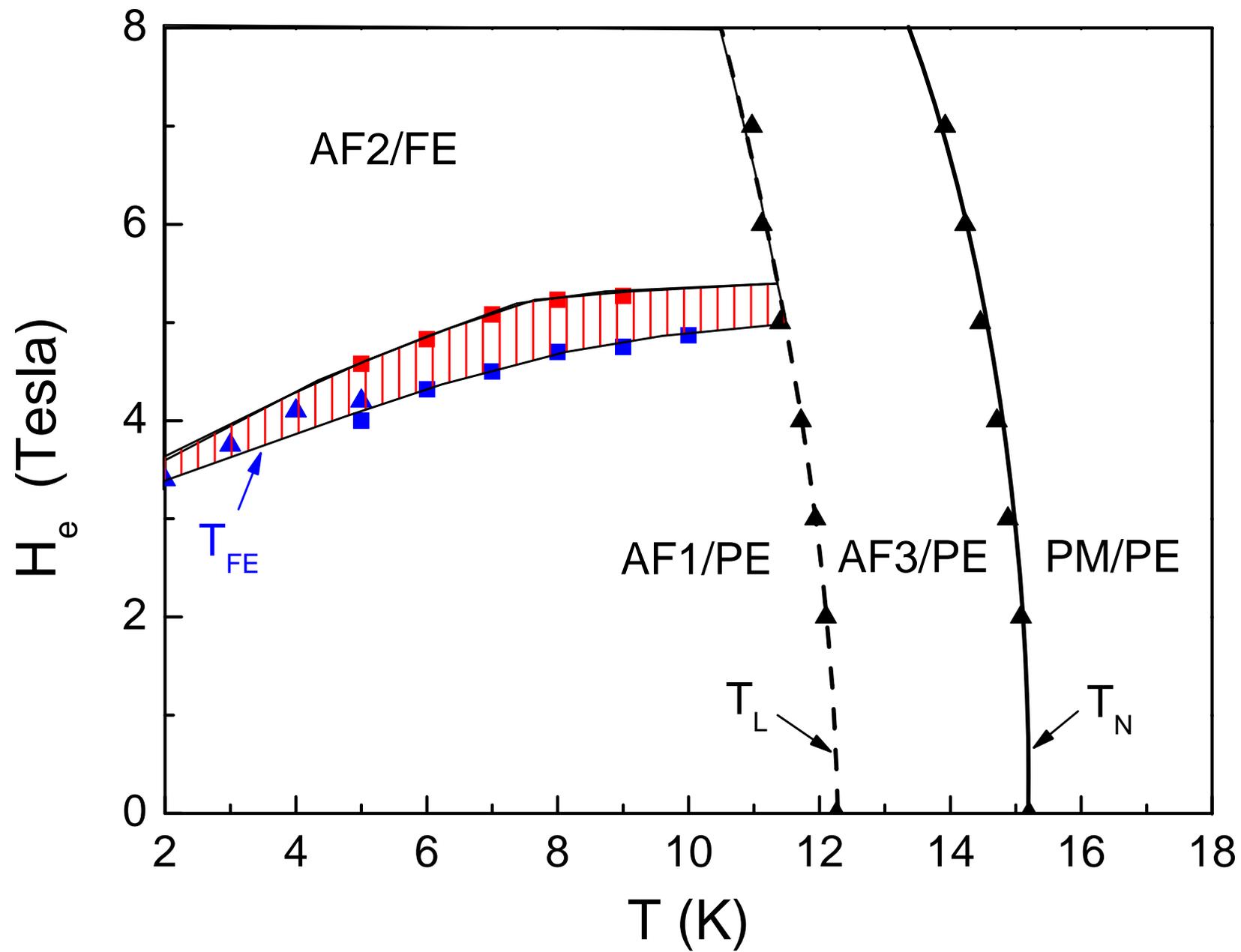
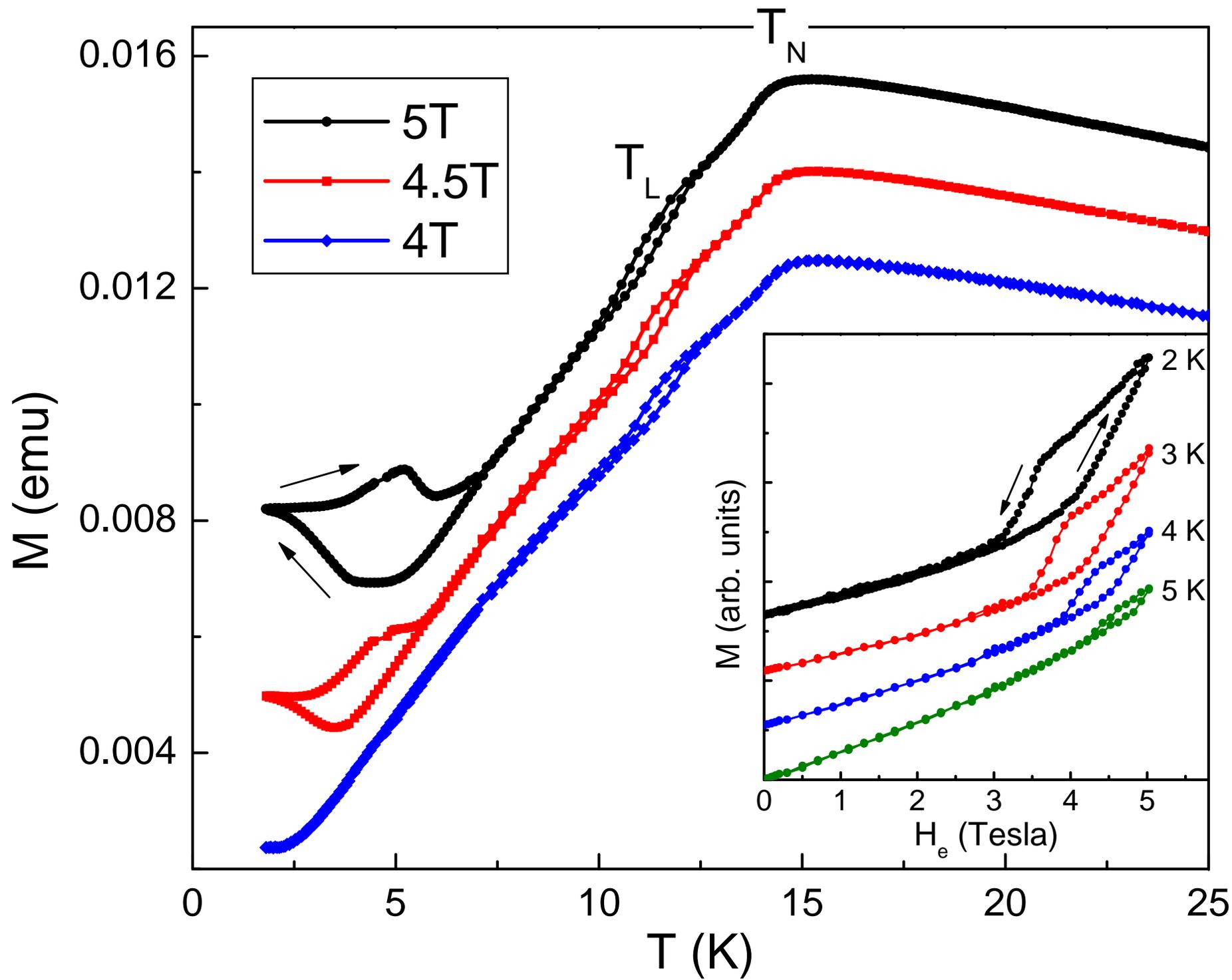
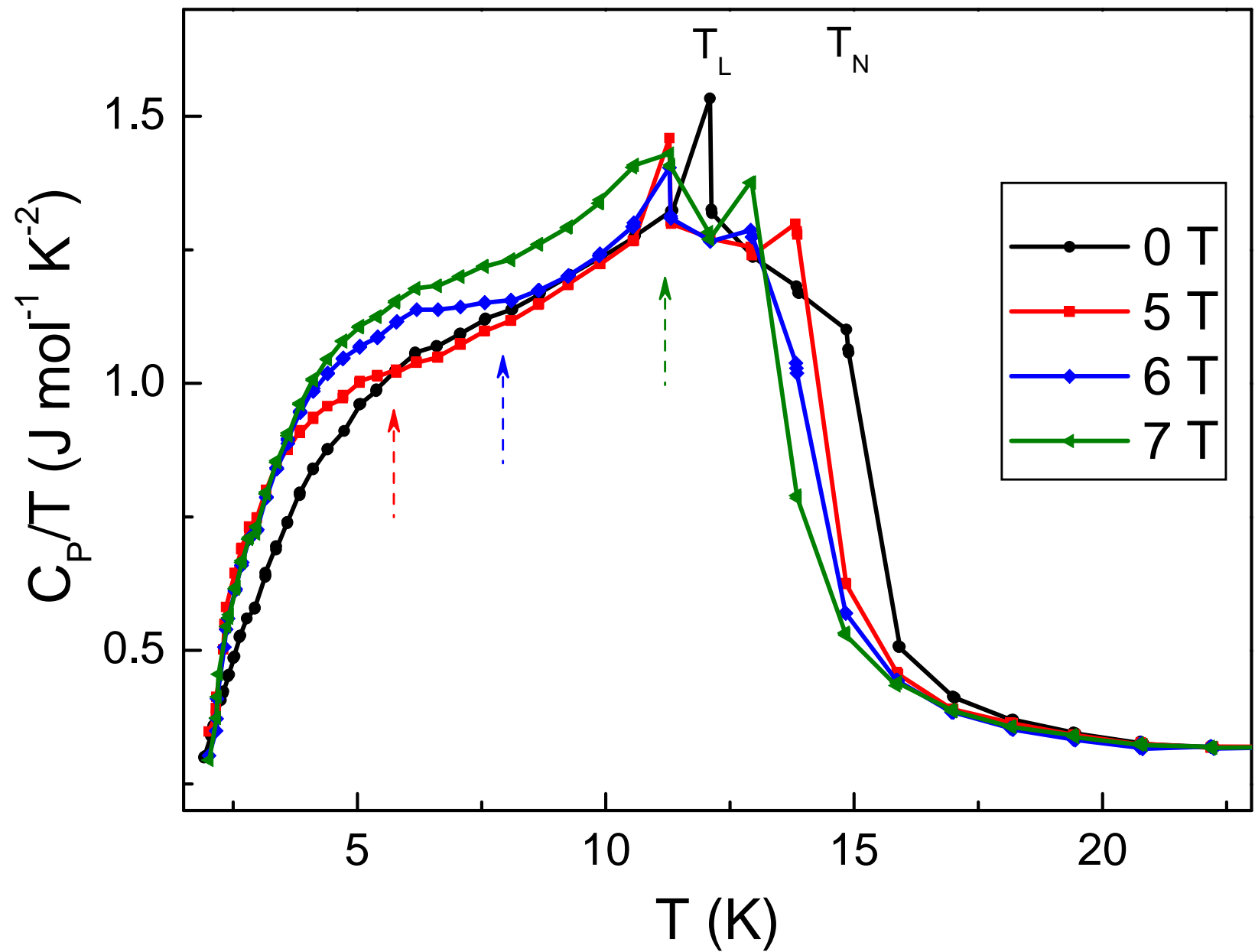
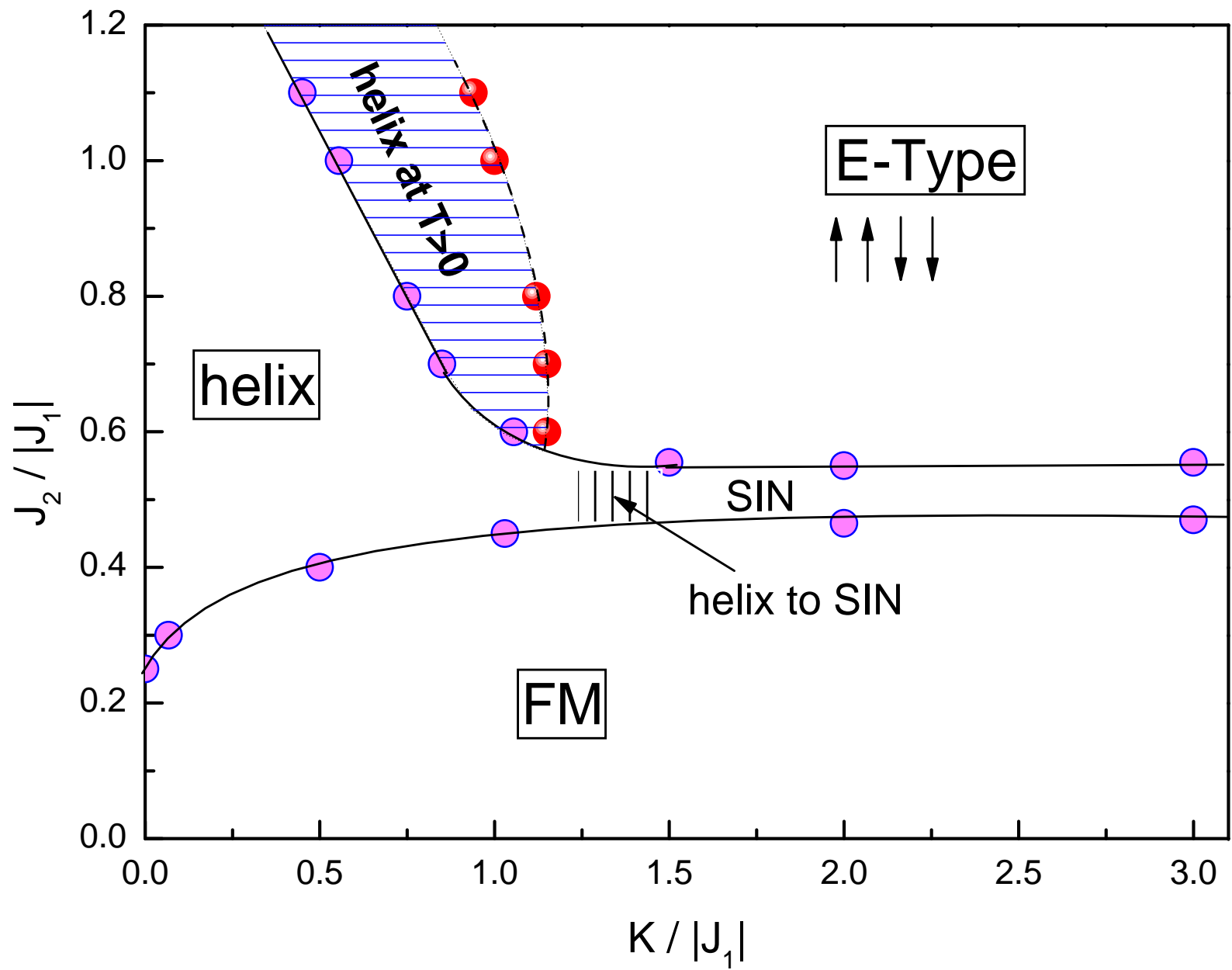
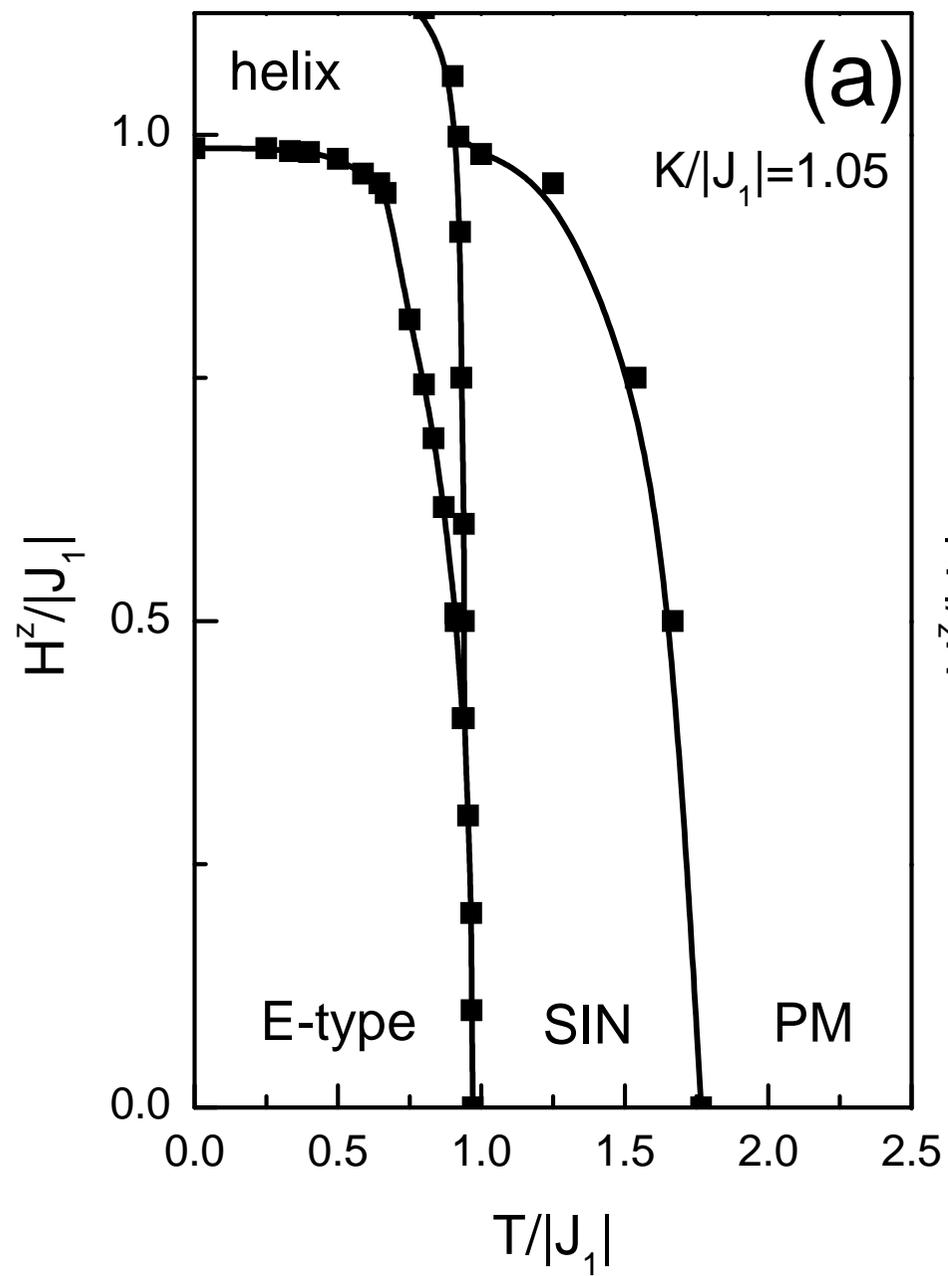
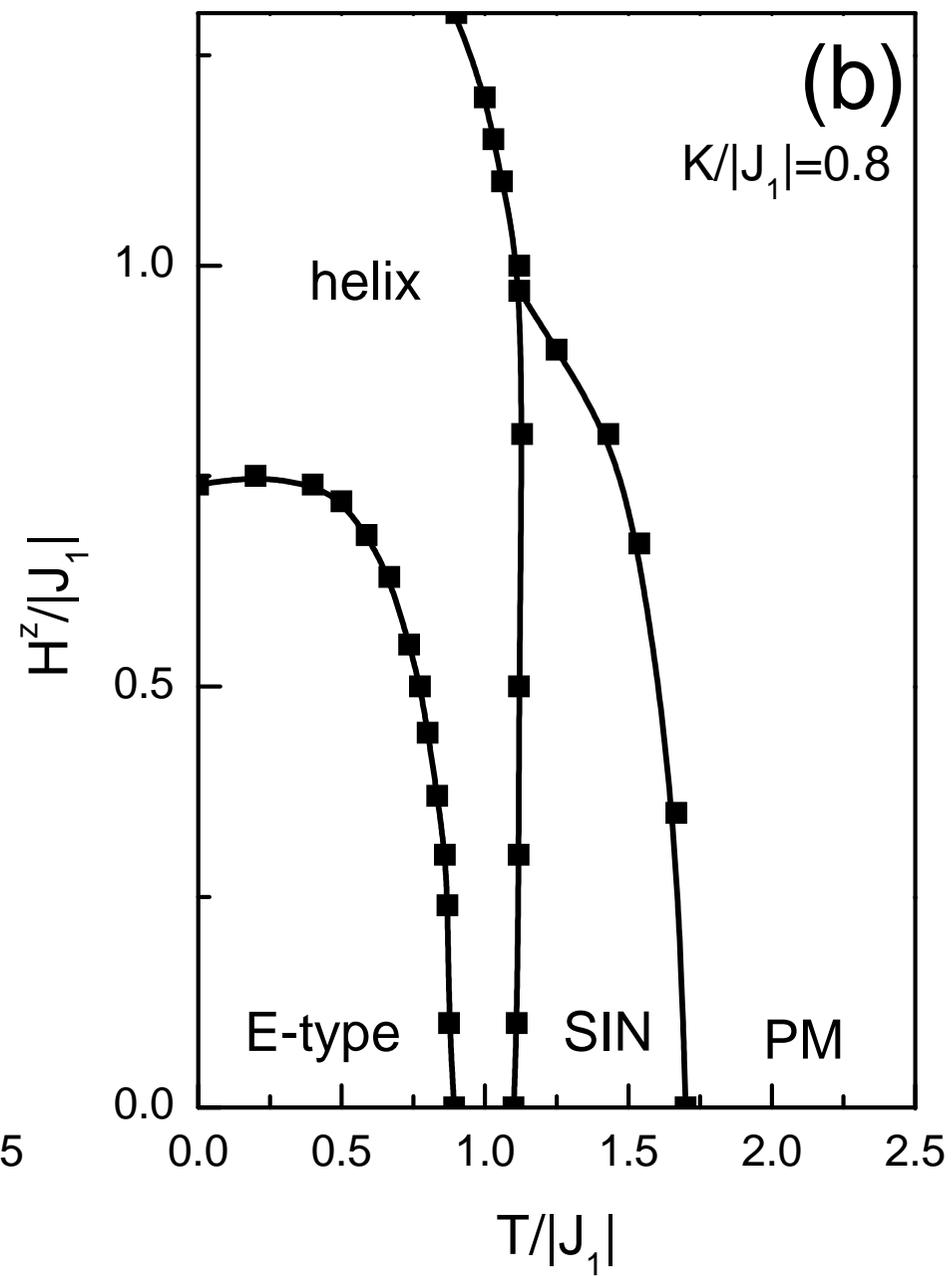
\end{document}